\documentclass[conference,10pt]{IEEEtran}

\usepackage[dvips]{graphicx}
\usepackage{float}
\usepackage{times}
\usepackage{amsmath}
\usepackage{amsfonts}
\usepackage{gensymb}
\usepackage{cite}
\usepackage{bm}
\usepackage{booktabs}
\usepackage{upgreek}
\usepackage{fancyhdr}

\begin{document}
%
\title{Direction Finding for a Mixture of Single-Transmission and Dual-Transmission Signals}


\author{\IEEEauthorblockN{Xiang Lan and Wei Liu }
	\IEEEauthorblockA{ 
		Department of Electronic and Electrical Engineering\\
		University of Sheffield, United Kingdom\\
{\it Email: \{xlan2, w.liu\}@sheffield.ac.uk}}
}

\maketitle
\thispagestyle{fancy}
\fancyhead{}
\lhead{}
\lfoot{978-1-7281-5341-4/19/\$31.00 \textcopyright 2019 IEEE}
\cfoot{}
\rfoot{}£¬

\begin{abstract}
Currently, most of existing research in direction of arrival (DOA) estimation is focused on single signal transmission (SST) based signal. However, to make full use of the degree of freedom provided by the system in the polarisation domain, the dual signal transmission (DST) model has been adopted more and more widely in wireless communications. In this work, a DOA estimation method for a mixture of SST and DST signals (referred to as the mixed signal transmission (MST) model) is proposed. To our best knowledge, this is the first time to study the DOA estimation problem for such an MST model. There are two steps in the proposed method, which deals with the two kinds of signals separately. The performance of the proposed method is compared with the Cram\'{e}r-Rao Bound (CRB) based on computer simulations.


\end{abstract}


%
\IEEEpeerreviewmaketitle

\section{Introduction}
Direction of arrival (DOA) estimation has been widely studied in recent years and many algorithms have been introduced to solve the problem, such as multiple signal classification (MUSIC) \cite{sch86}, estimation of signal parameters via rotational invariance techniques (ESPRIT) \cite{roy89} and those based on sparsity or compressive sensing (CS) \cite{kim12,liu15e,shen16}. In its early time, most research on DOA estimation was based on omnidirectional antennas, ignoring the polarization information of impinging signals. To consider the polarization information,  electromagnetic (EM) vector sensor arrays were proposed to jointly estimate the DOA and  polarization information \cite{compton81,li1993,nehorai1994,wong01,miron2006,xin08,liu13n}. The traditional algorithms can be extended to solve the joint DOA and polarization estimation problem \cite{xu04,zhang2014b,lan17,shuai09}. However, in their models, for each direction, it is assumed either explicitly or implicitly that there is only one signal impinging upon the array; in other words, each source only emits one single signal with specific direction and polarization and we refer to such a system as a single signal transmission (SST) system.

To make full use of the degree of freedom provided by a vector sensor array, two separate signals could be transmitted simultaneously from each source, and this is refer to the dual signal transmission (DST) model \cite{nehorai98}. For a DST signal, the two sub-signals have the same DOA but different polarizations. One DST example is to use two orthogonal linearly polarized signals with amplitude or phase modulation \cite{liu2017,lan2017,qureshi18}. However, there has rarely been any research reported on estimating the DOAs of DST signals. Instinctively, we could consider a DST signal as two independent SST signals and estimate their DOAs one by one. However, as we will see later, a direct application of the traditional DOA estimation methods such as the subspace-based ones may not work as expected for DST signals and a new approach is needed.

In this work, based on a uniform linear tripole sensor array, we first try to extend the classic MUSIC algorithm in a straightforward way to the 4-D case to find the parameters of a mixture of impinging SST and DST signals, which is referred to as the mixed signal transmission (MST) model. As analysed later, due to inherent physical property of signal polarisation and array structure, we can only find the DOA and polarisation parameters of SST signals and for the DST signals, it fails due to an ambiguity problem with their estimation. As a solution and also to reduce the complexity of the 4-D search process of the extended MUSIC algorithm and exploit the additional information provided by DST signals, i.e. the two sub-signals of each DST signal share the same DOA, a two-step algorithm is proposed. In this solution, the DOA and polarisation information of SST signals are found first by a rank-reduction algorithm (referred to as the SST estimator) and then the DOA information of the DST signals is estimated by a specifically designed estimator (referred to as the DST estimator). To our best knowledge, this is the first time to study the DOA estimation problem for such an MST model with a successful solution.

This paper is structured as follows. The MST signal model is introduced in Section II. In Section III, the traditional subspace based estimator is extended to the 4-D case, followed by the two-step method associated with SST and DST estimators. Finally, conclusions are drawn in Section VI.

\section{Signal Models}\label{sec:QM}
\begin{figure}
  \centering
  \includegraphics[width=0.24\textwidth]{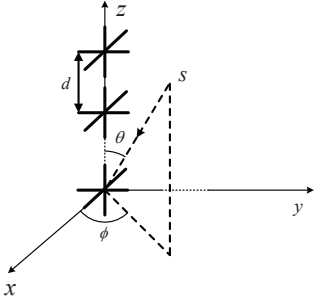}
  \caption{Structure of a uniform linear tripole array.}\label{fig:01}
\end{figure}

Assume that there are $M_1$ SST and $M_2$ DST narrowband non-linearly polarized sources impinging on a uniform linear array (ULA) with $N$ tripole sensors from the far field as shown in Fig. \ref{fig:01}. Each SST source generates one signal $\emph{s}_{m}(t),m=1, \cdots, M_1,$ and each DST source generates two sub-signals $\emph{s}_{M_1+2m-1}(t)$ and $\emph{s}_{M_1+2m}(t),m=1, \cdots, M_2,$ with the same elevation-azimuth angle $(\theta,\phi)$ but different polarization $(\gamma,\eta)$, where $\gamma$, $\eta$ denote the polarization auxiliary angle and the polarization phase difference, respectively.

For convenience, the parameters of the DST signal $\emph{s}_{M_1+2m-1}$ and $\emph{s}_{M_1+2m}$ are denoted by $(\theta_{M_1+2m-1},\phi_{M_1+2m-1},\gamma_{M_1+2m-1},\eta_{M_1+2m-1})$ and $(\theta_{M_1+2m},\phi_{M_1+2m},\gamma_{M_1+2m},\eta_{M_1+2m})$, respectively, with  $\theta_{M_1+2m-1} = \theta_{M_1+2m}$ and  $\phi_{M_1+2m-1} = \phi_{M_1+2m}$
%

In discrete form, the received SST signals of a single tripole sensor at the k-th time instant can be denoted by a $3\times 1$ vector $\emph{\textbf{x}}_s[k]$ (noise-free)
\begin{eqnarray}
    \emph{\textbf{x}}_s[k] =\sum_{m=1}^{M_1}\emph{\textbf{p}}_{m}s_{m}[k]
\end{eqnarray}
where $\emph{\textbf{p}}_m$ is the SST angular-polarization vector given by
\begin{eqnarray}
    \emph{\textbf{p}}_{m} &=&
                            \left[
                                \begin{matrix}
                                    \cos\theta_m\cos\phi_m & -\sin\phi_m \\
                                    \cos\theta_m\sin\phi_m & \cos\phi_m \\
                                    -\sin\theta_m & 0
                                \end{matrix}
                            \right]
                            \left[
                                \begin{matrix}
                                    \sin\gamma_{m} e^{j\eta_{m}} \\
                                    \cos\gamma_{m}
                                 \end{matrix}
                            \right] \nonumber\\
                         &=& \mathbf{\Omega}_m\cdot\emph{\textbf{g}}_{m}
                         \label{eq:polarisation}
\end{eqnarray}
In the above equation, $\mathbf{\Omega}_m$ denotes the angular matrix associated with DOA parameters $\theta$ and $\phi$, and $\emph{\textbf{g}}_m$ is the polarization vector including polarization parameters $\gamma$ and $\eta$.

The DST signals collected by a single tripole sensor can be considered as the sum of all $2M_2$ sub-signals, where each sub-signal can be viewed as a SST signal. Hence, the received DST signals are in the form
\begin{eqnarray}
    \emph{\textbf{x}}_d[k] &=&\sum_{m=M_1+1}^{M_1+2M_2}\emph{\textbf{p}}_{m}s_{m}[k]
    \label{eq:xd}
\end{eqnarray}
Considering a pair of sub-signals as a single composite DST signal, we can use a $2\times 1$ vector $\emph{\textbf{s}}_m$ to denote the m-th DST signal corresponding to the pair of sub-signals $s_{M_1+2m-1}$ and $s_{M_1+2m}$, which is defined by
\begin{eqnarray}
    \emph{\textbf{s}}_m[k] =\left[
                                \begin{matrix}
                                    s_{M_1+2m-1}[k] \\
                                    s_{M_1+2m}[k]
                                 \end{matrix}
                            \right]
\end{eqnarray}
Then, (\ref{eq:xd}) can be transformed to
\begin{eqnarray}
    \emph{\textbf{x}}_d[k] &=&\sum_{m=1}^{M_2}[\emph{\textbf{p}}_{M_1+2m-1} \quad \emph{\textbf{p}}_{M_1+2m-1}]
    \left[
                                \begin{matrix}
                                    s_{M_1+2m-1}[k] \\
                                    s_{M_1+2m}[k]
                                 \end{matrix}
                            \right]\nonumber\\
                            &=&\sum_{m=1}^{M_2}\emph{\textbf{P}}_m\emph{\textbf{s}}_m[k]
\end{eqnarray}
where $\emph{\textbf{P}}_m$ is the angular-polarization matrix for DST signals
\begin{eqnarray}
    \emph{\textbf{P}}_m &=&[\emph{\textbf{p}}_{M_1+2m-1} \quad \emph{\textbf{p}}_{M_1+2m-1}]\nonumber\\
    &=&[\mathbf{\Omega}_{M_1+2m-1} \emph{\textbf{g}}_{M_1+2m-1}\quad \mathbf{\Omega}_{M_1+2m}\emph{\textbf{g}}_{M_1+2m}]
\end{eqnarray}
Note that the two sub-signals of the same DST signal share the same angular matrix, and here we use $\mathbf{\Xi}_m$ to represent the common angular matrix of the m-th DST signal, i.e.
\begin{eqnarray}
    \mathbf{\Xi}_m=\mathbf{\Omega}_{M_1+2m-1} = \mathbf{\Omega}_{M_1+2m}
\end{eqnarray}
We use $\emph{\textbf{G}}_{m}$ to denote the polarization matrix of the m-th DST signal, defined as $\emph{\textbf{G}}_m=[\emph{\textbf{g}}_{M_1+2m-1}\quad \emph{\textbf{g}}_{M_1+2m}]$. Then, we have $\emph{\textbf{P}}_m = \mathbf{\Xi}_m\emph{\textbf{G}}_m$.

The total received signal $\emph{\textbf{x}}[k]$ is given by
\begin{equation}
    \emph{\textbf{x}}[k]=\sum_{m=1}^{M_1}\emph{\textbf{p}}_{m}s_{m}[k]
                +\sum_{m=1}^{M_2}\emph{\textbf{P}}_m\emph{\textbf{s}}_m[k]
\end{equation}

Now consider the whole array system. The steering vector $\emph{\textbf{a}}_m$ is defined by
\begin{eqnarray}
    \emph{\textbf{a}}_m=[1,e^{-j\tau \sin\theta_m},...,e^{-j(N-1)\tau \sin\theta_m}]^T
    \label{eq:steer}
\end{eqnarray}
where $\tau$ is the phase shift between two adjacent sensors.

With noise vector $\emph{\textbf{n}}[k]$, the array output $\emph{\textbf{y}}[k]$ at the k-th snapshot is either given by \cite{lan17}
\begin{eqnarray}
    \emph{\textbf{y}}[k] = \sum_{m=1}^{M_1+2M_2}\emph{\textbf{a}}_m\otimes\emph{\textbf{p}}_{m}\cdot s_{m}[k]+\emph{\textbf{n}}[k]
    \label{eq:joint2}
\end{eqnarray}
or
\begin{eqnarray}
    \emph{\textbf{y}}[k] &=& \sum_{m=1}^{M_1}\emph{\textbf{a}}_m\otimes\emph{\textbf{p}}_{m}\cdot s_{m}[k]\nonumber\\
    &+&\sum_{m=1}^{M_2}\emph{\textbf{a}}_{M_1+2m-1}\otimes\emph{\textbf{P}}_m\cdot \emph{\textbf{s}}_{m}[k]
    +\emph{\textbf{n}}[k]
    \label{eq:joint3}
\end{eqnarray}
Note that in the DST signal part, $\emph{\textbf{a}}_{M_1+2m-1}=\emph{\textbf{a}}_{M_1+2m}$.

Define $\emph{\textbf{q}}_{m}$ as the direction-polarization joint steering vector for SST signals, with
\begin{eqnarray}
    \emph{\textbf{q}}_m = \emph{\textbf{a}}_m\otimes\emph{\textbf{p}}_{m}
    \label{eq:qm1}
\end{eqnarray}
and $\emph{\textbf{Q}}_{m}$ as the DST joint steering matrix, given by
\begin{eqnarray}
    \emph{\textbf{Q}}_m &=& \emph{\textbf{a}}_{M_1+2m-1}\otimes\emph{\textbf{P}}_m\nonumber\\
                        &=& \emph{\textbf{a}}_{M_1+2m-1}\otimes[\emph{\textbf{p}}_{M_1+2m-1} \quad \emph{\textbf{p}}_{M_1+2m-1}]\nonumber\\
                        &=& [\emph{\textbf{q}}_{M_1+2m-1}\quad \emph{\textbf{q}}_{M_1+2m}]
    \label{eq:qm}
\end{eqnarray}
Then, (\ref{eq:joint3}) is simplified to
\begin{eqnarray}
    \emph{\textbf{y}}[k] &=& \sum_{m=1}^{M_1}\emph{\textbf{q}}_m\cdot s_{m}[k]
    +\sum_{m=1}^{M_2}\emph{\textbf{Q}}_m\cdot \emph{\textbf{s}}_{m}[k]
    +\emph{\textbf{n}}[k]
\end{eqnarray}

\section{Proposed Estimators}\label{sec:Without}
In this section, we first try to extend the classic MUSIC algorithm to the 4-D case to show that it will not work for DST signals, and then propose a two-step algorithm as a solution.


\subsection{Extension of the Traditional MUSIC Estimator to 4-D}
A traditional DOA estimator considers all incoming signals as individual SST signals, i.e. the given $M_1$ SST signals and $M_2$ DST signals will be considered as $M_1+2M_2$ SST signals.

For the general MST model, assume $M_1+2M_2< 3N$. With eigen-decomposition, the covariance matrix $\emph{\textbf{R}}=E\{\emph{\textbf{y}}[k]\emph{\textbf{y}}[k]^H\}$ can be expressed as
\begin{eqnarray}
    \emph{\textbf{R}} = \sum_{n=1}^{3N}\lambda_n\emph{\textbf{u}}_n\emph{\textbf{u}}_n^{H}
    \label{eq:eig}
\end{eqnarray}
where $\lambda_n$ is the $n$-th eigenvalue and $\emph{\textbf{u}}_n$ is the associated eigenvector. After sorting the $3N$ eigenvalues in descending order, the eigenvectors $\emph{\textbf{u}}_{1},...,\emph{\textbf{u}}_{M_1+2M_2}$ form the signal subspace $\emph{\textbf{U}}_s$, while $\emph{\textbf{u}}_{M_1+2M_2+1},...,\emph{\textbf{u}}_{3N}$ form the noise subspace $\emph{\textbf{U}}_n$. Both the SST joint steering vector $\emph{\textbf{q}}_m$ and the DST joint steering matrix $\emph{\textbf{Q}}_m$ are orthogonal to the noise subspace $\emph{\textbf{U}}_n$.

Then, we have
\begin{eqnarray}
    \emph{\textbf{U}}_n^{H}\emph{\textbf{a}}_m \otimes (\mathbf{\Omega}_{m}\emph{\textbf{g}}_m)=\textbf{0}
\end{eqnarray}

The DOA and polarization parameters could be found by the peaks of the following cost function through a 4-D search.
\begin{eqnarray}
    F(\theta,\phi,\gamma,\eta)=\frac{1}{\emph{\textbf{q}}_{m}^{H}
    \emph{\textbf{U}}_n\emph{\textbf{U}}_n^{H}\emph{\textbf{q}}_{m}}
\end{eqnarray}

However, as shown in the following, there is an ambiguity problem with both DOA and polarisation of DST signals, which can not be obtained by the subspace based method.

Firstly, the ambiguity problem associated with the polarisation parameters is analysed. Suppose that the two sub-signals of the DST signal comes from $(\theta,\phi,\gamma_1,\eta_1)$ and $(\theta,\phi,\gamma_2,\eta_2)$, and they are also considered as two separate SST signals. $\emph{\textbf{a}}$ is used to denote the common steering vector of the two sub-signals. $\emph{\textbf{p}}_1$ and $\emph{\textbf{p}}_2$ are used to denote their angular-polarisation vectors based on distinct polarisation parameters. According to (\ref{eq:polarisation}), it can be obtained that
\begin{eqnarray}
    \emph{\textbf{p}}_1 = \mathbf{\Omega}\emph{\textbf{g}}_{1},
    \emph{\textbf{p}}_2 = \mathbf{\Omega}\emph{\textbf{g}}_{2}
    \label{eq:g1}
\end{eqnarray}
Consider a non-existing signal from the same DOA of the sub-signals above $(\theta,\phi)$ with an arbitrary polarisation $(\gamma_3,\eta_3)$ different from $(\gamma_1,\eta_1)$ and $(\gamma_2,\eta_2)$. The angular-polarisation vector $\emph{\textbf{p}}_3$ can be denoted as
\begin{eqnarray}
    \emph{\textbf{p}}_3 = \mathbf{\Omega}\emph{\textbf{g}}_{3}
    \label{eq:g3}
\end{eqnarray}
From (\ref{eq:polarisation}), it can be learned that $\emph{\textbf{g}}_{m}, m\in[1,3]$, is a column vector with two elements and $\mathbf{\Omega}$ is a matrix with two columns. Then (\ref{eq:g1}) and (\ref{eq:g3}) can be changed to
\begin{eqnarray}
    \emph{\textbf{p}}_m = g_{m1}\mathbf{\upomega}_1 +g_{m2}\mathbf{\upomega}_2
    \label{eq:lin}
\end{eqnarray}
where $\mathbf{\upomega}_1$ and $\mathbf{\upomega}_2$ denote the first and second column vector of $\mathbf{\Omega}$. $g_{m1}$ and $g_{m2}$ are the first and second elements in $\emph{\textbf{g}}_{m}$, respectively. As $\mathbf{\upomega}_1$ is not in parallel with $\mathbf{\upomega}_2$ from (\ref{eq:polarisation}), (\ref{eq:lin}) indicates that $\emph{\textbf{p}}_1$, $\emph{\textbf{p}}_2$ and $\emph{\textbf{p}}_3$ are three vectors in the same two-dimensional space determined by $\mathbf{\upomega}_1$ and $\mathbf{\upomega}_2$. That means there exists a linear relationship among $\emph{\textbf{p}}_1$, $\emph{\textbf{p}}_2$ and $\emph{\textbf{p}}_3$, which can be denoted by
\begin{eqnarray}
    \emph{\textbf{p}}_3 = \lambda_1\emph{\textbf{p}}_1+\lambda_2\emph{\textbf{p}}_2
    \Rightarrow \emph{\textbf{q}}_3 = \lambda_1\emph{\textbf{q}}_1+\lambda_2\emph{\textbf{q}}_2
    \label{eq:q123}
\end{eqnarray}
where $\lambda_1$ and $\lambda_2$ are constants.

When the estimator is applied to a DST signal, the noise subspace will be orthogonal to the joint steering vectors of both sub-signals, where
\begin{eqnarray}
    \emph{\textbf{U}}_n^{H}\emph{\textbf{a}} \otimes \emph{\textbf{p}}_{1}=\textbf{0},
    \emph{\textbf{U}}_n^{H}\emph{\textbf{a}} \otimes \emph{\textbf{p}}_{2}=\textbf{0}
\end{eqnarray}
Then, we have
\begin{eqnarray}
    \emph{\textbf{U}}_n^{H}\emph{\textbf{a}} \otimes \emph{\textbf{p}}_{3}
    =\emph{\textbf{U}}_n^{H}\emph{\textbf{a}} \otimes (\lambda_1\emph{\textbf{p}}_1+\lambda_2\emph{\textbf{p}}_2)
    =\textbf{0}
\end{eqnarray}

As a result, $F(\theta_,\phi_,\gamma_{3},\eta_{3})$ will be recognised as a peak in the spectrum and wrongly identified as the parameters of a non-existing source. This means the algorithm fails when trying to estimate the polarization of DST signals. Note that this is an inherent limitation to the DST signals and there is no way to identify their polarisation parameters.

Next, we give an analysis to the ambiguity problem associated with the DOA of DST signals. (\ref{eq:q123}) shows that by 4-D MUSIC, the DST signal's direction with arbitrary polarisation will be recognised as a false peak. The `arbitrary polarisation' includes a special polarisation: the linear polarisation. As introduced in \cite{ho98}, there are infinite number of ambiguity steering vectors in parallel with a linearly polarised signal, where the ambiguity directions are in linear polarisation as well. From (\ref{eq:polarisation}), the angular-polarisation vector $\emph{\textbf{p}}_{m}$ can be viewed as an arbitrary vector in a 2-D space built by the two column vectors of $\mathbf{\Omega}_{m}$, where the elements in $\emph{\textbf{g}}_{m}$ denote the weights of the vectors to indicate how these two column vectors form $\emph{\textbf{p}}_{m}$. If a signal $s_1$ is linearly polarised, it means that the angular-polarisation vector $\emph{\textbf{p}}_{1}$ is real-valued. As the two column vectors in $\mathbf{\Omega}_{1}$ are both real-valued, the intersection vector between $\mathbf{\Omega}_{1}$ and another different 2-D space $\mathbf{\Omega}_{2}$ is also real-valued. Consequently, it is possible to locate the angular-polarisation vector $\emph{\textbf{p}}_{1}$ as the intersection vector between $\mathbf{\Omega}_{1}$ and $\mathbf{\Omega}_{2}$. This means in the 2-D space $\mathbf{\Omega}_{2}$, there exists another angular-polarisation vector $\emph{\textbf{p}}_{2}$ in parallel with $\emph{\textbf{p}}_{1}$. If $\emph{\textbf{a}}_{1}$ is also in parallel with $\emph{\textbf{a}}_{2}$, for example, $\theta_1=\theta_2$ while $\phi_1 \neq \phi_2$, $\emph{\textbf{q}}_{2}$ will be in parallel with $\emph{\textbf{q}}_{1}$ and the parameters in $\emph{\textbf{q}}_{2}$ will be recognised as a false peak.

For the MST scenario, although the 4-D search algorithm cannot identify the DST signals, it works for SST signals. However, an obvious problem is its very high computational complexity. In the next subsection, a low-complexity two-step algorithm is proposed, which estimates the DOAs of SST and DST signals separately.

\subsection{The Proposed Two-Step Method}

In the proposed method, the first step is to apply a newly proposed SST estimator to obtain the DOA and polarization of SST signals, while the second step is to apply a specifically designed DST estimator to find the DOA of DST signals.

In the first step, we focus on the SST signals. By exploiting the orthogonality between the joint steering vector $\emph{\textbf{q}}_m$ and the noise subspace $\emph{\textbf{U}}_n$ and with $\emph{\textbf{B}}_m=\emph{\textbf{a}}_m \otimes \mathbf{\Omega}_{m}$, we have
\begin{eqnarray}
    \textbf{0}&=&\emph{\textbf{U}}_n^{H}[\emph{\textbf{a}}_m \otimes (\mathbf{\Omega}_{m}\emph{\textbf{g}}_m)]\nonumber\\
    &=&\emph{\textbf{U}}_n^{H}[(\emph{\textbf{a}}_m \otimes \mathbf{\Omega}_{m}) \emph{\textbf{g}}_m]
    =[\emph{\textbf{U}}_n^{H}\emph{\textbf{B}}_m]\emph{\textbf{g}}_m
    \label{tran}
\end{eqnarray}

For SST signals, there is only one polarization vector $\emph{\textbf{g}}_m$ from a specific direction $(\theta_m,\phi_m)$ satisfing $[\emph{\textbf{U}}_n^{H}\emph{\textbf{B}}_m]\emph{\textbf{g}}_m=\textbf{0}$ and (\ref{tran}) indicates that the column rank of $\emph{\textbf{U}}_n^{H}\emph{\textbf{B}}_m$ equals to 1. Notice that $\emph{\textbf{U}}_n^{H}\emph{\textbf{B}}_m$ is a $(3N-M_1-2M_2)\times 2$ matrix. By multiplying its Hermitian transpose on the right side, the product matrix is a $2 \times 2$ matrix with rank 1, i.e.,
\begin{eqnarray}
    rank\{\emph{\textbf{B}}_m^{H}
    \emph{\textbf{U}}_n\emph{\textbf{U}}_n^{H}\emph{\textbf{B}}_m \}=1
    \label{eq:r1}
\end{eqnarray}
As the matrix is not of full rank, we have
\begin{eqnarray}
    det\{\emph{\textbf{B}}_m^{H}\emph{\textbf{U}}_n\emph{\textbf{U}}_n^{H}\emph{\textbf{B}}_m\}=0
    \label{det}
\end{eqnarray}
where $det\{\}$ represents the determinant of the matrix.
By taking the inverse of (\ref{det}), a DOA estimator for SST signals is given by
\begin{eqnarray}
    F_1(\theta_m,\phi_m)=\frac{1}{det\{\emph{\textbf{B}}_m^{H}
    \emph{\textbf{U}}_n\emph{\textbf{U}}_n^{H}\emph{\textbf{B}}_m \}}
    \label{eq:e1}
\end{eqnarray}
With the DOA information obtained, the polarization parameters can then be estimated through another 2-D search using (\ref{tran}). Besides, the first step will also detect the desired DOA angles of DST signals but with an infinite number of ambiguity directions. In the next step, the DOA of DST signals will be extracted from these results.

Since a DST signal $\emph{\textbf{s}}_m$ consists of two sub-signals $s_{M_1+2m-1}$ and $s_{M_1+2m}$ with different polarizations, we have
\begin{eqnarray}
    &&[\emph{\textbf{U}}_n^{H}\emph{\textbf{B}}_{M_1+2m-1}]\emph{\textbf{g}}_{M_1+2m-1} = \textbf{0}\nonumber\\
    &&[\emph{\textbf{U}}_n^{H}\emph{\textbf{B}}_{M_1+2m}]\emph{\textbf{g}}_{M_1+2m} = \textbf{0}
\end{eqnarray}
Since $\emph{\textbf{B}}_{M_1+2m-1}=\emph{\textbf{B}}_{M_1+2m}$, $\emph{\textbf{g}}_{M_1+2m-1}$ and $\emph{\textbf{g}}_{M_1+2m}$ are two distinct null vectors for $\emph{\textbf{U}}_n^{H}\emph{\textbf{b}}_{M_1+2m-1}$, which means $\emph{\textbf{U}}_n^{H}\emph{\textbf{B}}_{M_1+2m-1}$ is a zero matrix. Hence, the following cost function can be used to estimate directions of DST signals
\begin{eqnarray}
    F_2(\theta_n,\phi_n)=\frac{1}{||\emph{\textbf{B}}_{m}^{H}\emph{\textbf{U}}_n
    \emph{\textbf{U}}_n^{H}\emph{\textbf{B}}_{m}||_2}
    \label{eq:e2}
\end{eqnarray}
where $||\cdot||_2$ denotes the $l_2$-norm of the vector.

When the above DST estimator is applied to a mixture of SST and DST signals, it  only selects directions with
\begin{eqnarray}
    rank\{\emph{\textbf{b}}_m^{H}\emph{\textbf{U}}_n
    \emph{\textbf{U}}_n^{H}\emph{\textbf{b}}_m\}=0
\end{eqnarray}
However, for SST signals, (\ref{eq:r1}) indicates that the rank of $\emph{\textbf{b}}_m^{H}\emph{\textbf{U}}_n
\emph{\textbf{U}}_n^{H}\emph{\textbf{b}}_m$ is 1 and therefore the DST estimator in (\ref{eq:e2}) will miss the SST signals.

A summary to the proposed two-step algorithm:
\begin{itemize}
  \item Calculate the noise subspace $\emph{\textbf{U}}_n$ by applying eigenvalue decomposition to the estimated covariance matrix $\hat{\emph{\textbf{R}}}$.
  \item Apply the SST estimator (\ref{eq:e1}) and find the DOAs of SST signals by 2-D search.
  \item Find the polarization parameters of SST signals using (\ref{tran}) by 2-D search if needed.
  \item Apply (\ref{eq:e2}) to estimate the DOAs of DST signals.
\end{itemize}

\section{Simulation Results}
In this section, simulations are performed based on a scenario with one SST signal and one DST signal impinging on the array from the far field.
A uniform linear tripole sensor array is employed with $M=5$ sensors and an inter-element distance $d$ equal to $\lambda/2$. The SST signal and each sub-signal of a DST signal have the same power $\sigma_s^2$. The SST signal comes from $(\theta_1,\phi_1,\gamma_1,\eta_1)=(20^\circ,20^\circ,50^\circ,10^\circ)$ and the DST signal comes from $(\theta_2,\phi_2,\gamma_{2},\eta_{2},\gamma_{3},\eta_{3})=(60^\circ,60^\circ,20^\circ,
50^\circ,70^\circ,-40^\circ)$. 



\begin{figure}
  \centering
  \includegraphics[width=0.48\textwidth]{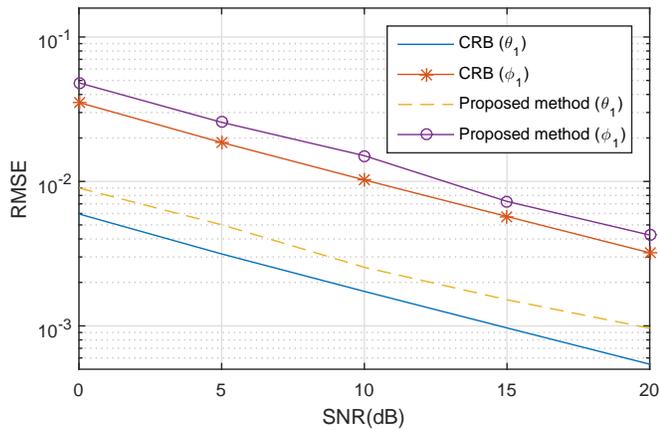}
  \caption{RMSE of SST DOA versus SNR.}
  \vspace{2ex}
  \label{fig:05}
\end{figure}

\begin{figure}
  \centering
  \vspace{1ex}
  \includegraphics[width=0.48\textwidth]{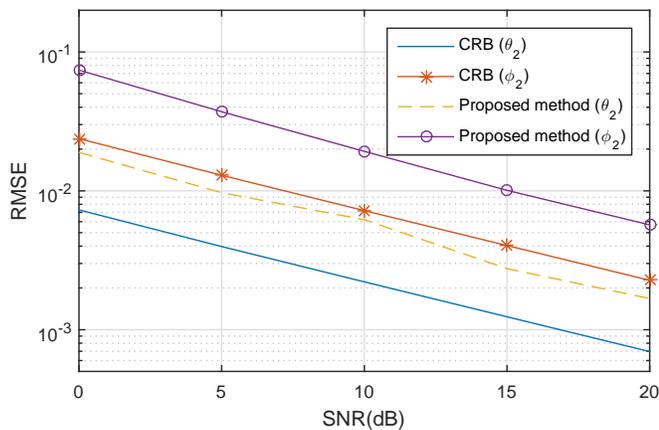}
  \caption{RMSE of DST DOA versus SNR.}
  \label{fig:07}
\end{figure}

The performance of the proposed method is compared with the Cram\'{e}r-Rao bound (CRB). The RMSE (root mean square error) of the azimuth-elevation angle $\theta,\phi$ is calculated by 200 Monte-Carlo trials. The number of snapshots is $K=100$ and the searching step size is $0.1\degree$.
As shown in Fig. \ref{fig:05} and Fig. \ref{fig:07}, with the increase of SNR, both RMSE and CRB of all four DOA parameters gradually decrease. Although the DST signal has a higher total SNR than the SST signal, the RMSE of SST signal's DOA ($\theta_1$ and $\phi_1$) is closer to the CRB than that of the DST signal ($\theta_2$ and $\phi_2$).

\section{Conclusion}

In this paper, the DOA estimation problem for a mixture of SST and DST signals has been studied based on a tripole linear array for the first time. Firstly, an analysis of the direct 4-D extension of the original MUSIC algorithm was provided to explain that the 4-D MUSIC algorithm has an ambiguity problem with mixed signals. To avoid the DOA ambiguity problem, a new subspace based DOA estimation method was proposed with two steps. The proposed method estimates the SST and DST signals' directions separately with two corresponding estimators, one for the SST signals and one for the DST signals. Simulation results showed that the proposed method was able to solve the estimation problem of mixed source signals effectively.


\bibliographystyle{IEEEtran}
\bibliography{mybib}

\end{document}